\documentclass[12pt,preprint]{aastex}
\input{epsf.sty}                        

\shorttitle{AGN Intrinsic Luminosity Function} \shortauthors{S.N.
Zhang}

\begin{document}

\title{A Single Intrinsic Luminosity Function for Both Type-I and Type-II Active Galactic Nuclei}

\author{Shuang Nan Zhang\altaffilmark{1,2,3,4}}
\affil{\altaffilmark{1}Physics Department and Center for
Astrophysics, Tsinghua University, Beijing, 100084, China
(zhangsn@tsinghua.edu.cn)} \affil{\altaffilmark{2}Physics
Department, University of Alabama in Huntsville, Huntsville,
AL35899, USA (zhangsn@uah.edu)} \affil{\altaffilmark{3}Space
Science Laboratory, NASA Marshall Space Flight Center, SD50,
Huntsville, AL35812, USA} \affil{\altaffilmark{4}Institute of High
Energy Physics, Chinese Academy of Sciences, Beijing, China}

\begin{abstract}
The luminous electromagnetic emission from distant active galactic
nuclei (AGNs) including quasars is believed to be powered by
accretion onto super-massive black holes (SMBHs).  In the standard
unification model for AGNs a dusty torus covers a significant
portion of the viewing angles to the accretion disk and the BH.
The system is classified as a type-I AGN if the accretion disk is
viewed through the opening part; otherwise it is called a type-II
AGN. Therefore the ratio of type-II to type-I AGNs serves as a
sensitive probe to the unification model. A surprising discovery
made from several large sky coverage and/or deep AGN surveys has
found a significant anti-correlation between the type-II fraction
and the observed X-ray luminosity between 2-10 keV. This suggests
two different luminosity functions for the two types of AGNs, thus
challenging the AGN unification model. However this observed
anti-correlation is a natural consequence of the AGN unification
model with only one intrinsic luminosity function if the
inclination angle effects of the X-ray emitting accretion disk are
taken into account. Thus the AGN unification model survived
another critical test.
\end{abstract}

 \maketitle

\section{Introduction}

Observationally type-I AGNs are seen to have soft X-ray spectra
(below about 10 keV) and both narrow and broad emission lines in
their optical spectra, in contrast to type-II AGNs with harder
X-ray spectra and only narrow optical emission lines
\cite{Antonucci}. These observations are naturally explained in
the standard unification model of AGNs \cite{Antonucci} in which
both the X-ray emission and the broad emission lines are produced
within the region very close to the BH; the dusty torus blocks
this region when viewed nearly edge-on (type-II AGNs) to the dusty
torus. Because the dusty torus absorbs the broad optical emission
lines almost completely and X-ray photons with lower energies
suffer more absorption than higher energy photons, the type-II
AGNs are observed to have harder X-ray spectra and do not show
obvious broad optical emissions lines. Evidence has been
accumulated from many different observations in infrared, optical
and X-ray bands in support of this unification model for the two
types of AGNs \cite{Antonucci}. Despite of these progresses, no
physically consistent model is currently available to account for
the formation and evolution of the dusty torus, which may provide
the crucial link between the galactic structure at larger scales
and the accretion disk which fuels the SMBHs.

\section{Correlation between X-ray luminosity and type-II AGN fraction}

Recently it has been found that the torus structure may be
different for AGNs with different X-ray luminosity, because the
fraction of type-II AGNs is anti-correlated with the observed
X-ray luminosity, e.g., found from combined {\it ASCA}, {\it
HEAO1} and {\it Chandra} surveys \cite{Ueda}, from combined {\it
Chandra} and {\it XMM-Newton} surveys \cite{Hasinger}, from
combined {\it ASCA} and {\it Chandra} surveys \cite{Steffen}, from
{\it RXTE} slew survey \cite{Sazonov}, and from a sample of PG
AGNs \cite{Wang}. Therefore the above unification scheme may need
modifications. It is proposed that the smaller type-II fraction
for more X-ray luminous AGNs may imply that the X-ray radiation is
blowing out the dusty torus, such that the opening angles for more
luminous AGNs become larger \cite{Ueda,Hasinger,Barger}. However
the dusty torus may also evolve by itself due to the dissipations
of collisions among the clouds inside the torus
\cite{Krolik,Wang2}. Despite of these progresses, the formation
and evolution of the dusty torus, which may have important
consequences for the formation and evolution of SMBHs and their
host galaxies, are still poorly understood.

However the observed anti-correlation between type-II AGN
fractions and X-ray luminosity may be naturally explained within
the standard AGN unification model if the planes of the accretion
disk and the torus are co-aligned and the X-ray emission is
produced mainly from the optically thick accretion disk. In this
case, type-II AGNs are viewed nearly edge-on to both the torus and
the accretion disk. Because a smaller X-ray flux is observed from
an edge-on disk due to the less projected area of the disk,
type-II AGNs appear to be less luminous than type-I AGNs for the
same intrinsic luminosity. The observed apparent X-ray luminosity
is reduced by a factor of $\cos(\theta)(1+2\cos(\theta))/3$, where
$\theta$ is the inclination angle of the accretion disk and
$\theta=90$ degrees for an edge-on disk; the factor of
$\cos(\theta)$ is due to the area-projection effect and the factor
of $ (1+2\cos(\theta))/3$ is due to the limb-darkening effect
\cite{Netzer} respectively (though our calculations show the
simple projection effect alone would produce almost identical
results). If AGNs are assumed being oriented randomly in the sky,
then the probability of seeing an AGN at an inclination angle
$\theta$ is proportional to $\sin(\theta)$. Therefore for a given
intrinsic luminosity of a group of AGNs, the observed apparent
luminosity follows a distribution proportional to
$f(x)=\sqrt{1-x^2}(1+2\sqrt{1-x^2})$, where $x=\sin(\theta)$ is
uniformly distributed between 0 and 1. Here we ignore all possible
inclination angle dependent relativistic effects which may change
both the observed X-ray flux and spectral shape if the emission
region is close to the BH \cite{Zhang}, because the present AGN
statistics does not require further refinement to this simple
model. Consequently the convolution between $f(x)$ and a given
intrinsic luminosity function produces the observed apparent
luminosity function, as shown in Figure 1.
\begin{figure}
\centerline{ \hbox{ \epsfxsize=6.0in \epsfbox{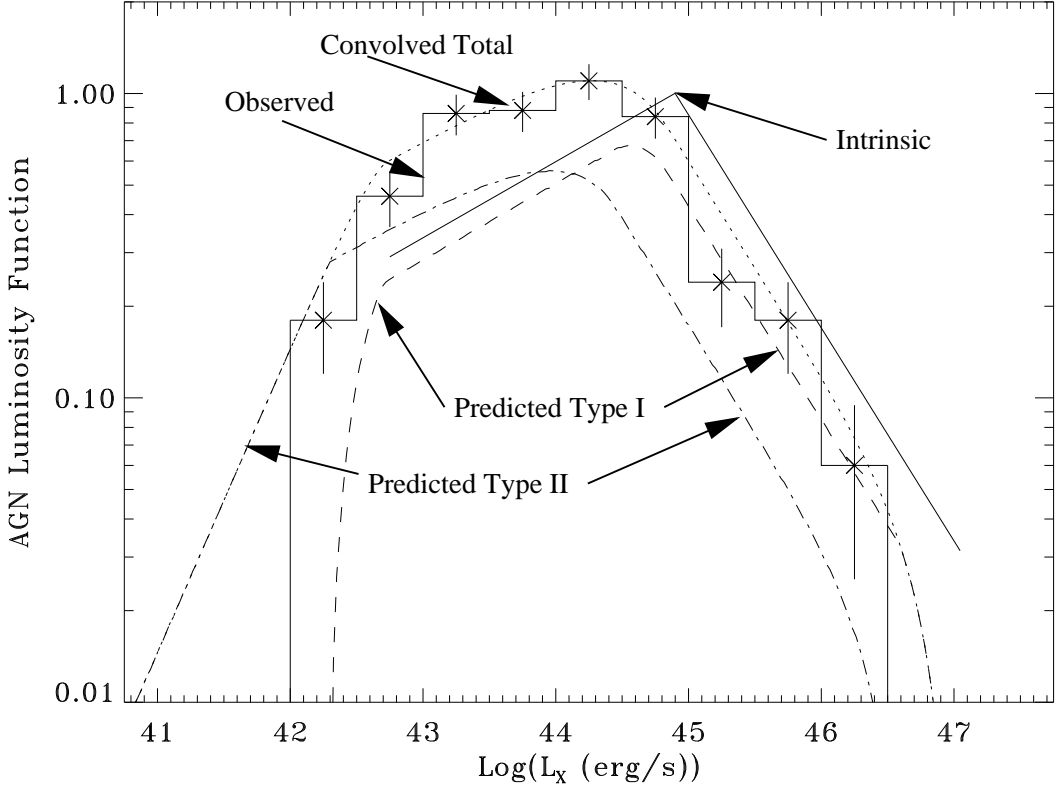} }}
  \caption{\small AGN luminosity function. The intrinsic luminosity function, referred
  to as the AGN luminosity before correcting for the inclination angle effects,
  is assumed of a broken power-law shape, i.e., $N\propto
L_{X}^{\alpha}$, where $L_{X}$ is in units of erg/s, $\alpha=0.25$
for $10^{42.75}<L_{X}\leq10^{44.9}$ and $\alpha=-0.7$ for
$10^{44.9}<L_{X}<10^{47}$; these parameters are determined by
matching the data with the model predictions. The observed
luminosity distribution (after absorption corrections) of AGNs
\cite{Ueda} agree with the predicted apparent luminosity defined
as $L_X=F_X4\pi D_{L}^{2}$, where $F_X$ is the observed X-ray flux
and $D_L$ is the luminosity distance of the AGN. The predicted
type-I and type-II AGN luminosity functions are also shown for
comparison, if AGNs with inclination angles greater than 68
degrees are classified as type-II AGNs. Clearly in the low
luminosity range type-II AGNs dominates, in contrast to the high
luminosity range where one finds mostly type-I AGNs.}\label{fig1}
\end{figure}
\begin{figure}
\centerline{ \hbox{ \epsfxsize=6.0in \epsfbox{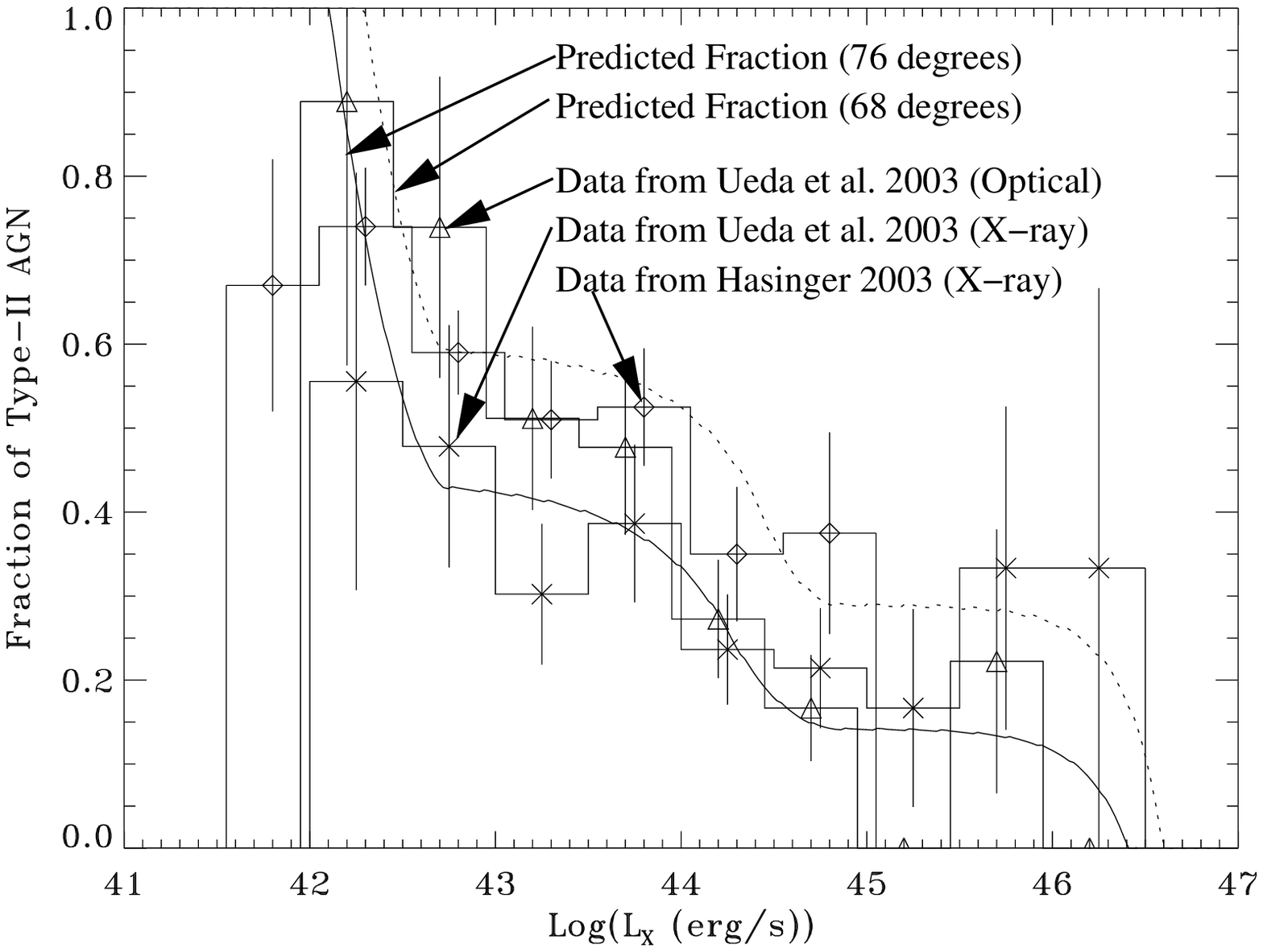} }}
  \caption{\small Type-II AGN fraction as function of the observed
apparent X-ray luminosity (after absorption corrections). The data
points shown by diamonds and triangles are shifted horizontally by
0.05 and -0.05 respectively for displaying clarity. Because the
three different groups of type-II AGNs, i.e., optical and X-ray
type-II AGNs from Ueda {\it et al.} (2003) and X-ray type-II AGNs
from Hasinger (2003) may have slightly different definitions in
terms of the dividing inclination angle between type-I and type-II
AGNs, we also show two different model predictions corresponding
to two critical inclination angles of 68 and 76 degrees
respectively. We did not include the model-fitted relation between
type-II AGN fraction and X-ray luminosity by Ueda {\it et al.}
(2003) because the relation contains only three values over the
entire luminosity range, lacking details for comparison to our
model predictions with several distinctive features; the general
trend of the three values is not significantly different from the
``raw" data points shown here.}\label{fig12}
\end{figure}

In Figure 1, we apply the above mentioned simple inclination angle
effects to the AGNs sample used by Ueda {\it et al.} (2003); all
data points are from Figure 4 of Ueda {\it et al.} (2003). A
simple broken-power law form of the intrinsic AGN luminosity
function is first assumed, in order to mimic the overall features
of the observed apparent luminosity function. We then convolve
between $f(x)$ and this intrinsic luminosity function to produce a
trial apparent luminosity function. By adjusting the parameters of
the assumed intrinsic luminosity function and compare each trial
apparent luminosity function, the best estimates for these
parameters are determined: $N\propto L_{X}^{\alpha}$, where
$L_{X}$ is in units of erg/s, $\alpha=0.25$ for
$10^{42.75}<L_{X}\leq10^{44.9}$ and $\alpha=-0.7$ for
$10^{44.9}<L_{X}<10^{47}$. It is clear that the observed apparent
X-ray luminosity function (after absorption corrections) is
significantly different from the assumed simple intrinsic
luminosity form. If AGNs with inclination angles greater than 68
degrees are classified as type-II AGNs, we also show the predicted
apparent luminosity functions for both type-I and type-II AGNs;
clearly these two luminosity functions are also drastically
different from each other.

In Figure 2, our model predicted type-II fraction as function of
the observed apparent X-ray luminosity is compared to the observed
anti-correlation. The data points are taken from Ueda {\it et al.}
(2003) (Figure 4) and Hasinger (2003) (Figure 6, left panel), as
indicated in Figure 2. The model predicted type-II AGN fraction is
calculated as the ratio between the predicted apparent type-II AGN
luminosity function and the observed apparent total luminosity
function, as shown in Figure 1. Because different samples of
type-II AGNs may have different dividing inclination angles, we
show our model predictions for two different dividing inclination
angles of 68 and 76 degrees respectively. Not only our model
re-produces the observed overall tendency of the anti-correlation,
several features of the observed anti-correlation also agree with
the model predictions well (albeit the limited statistics in the
data), e.g., the rapidly declining region for
$10^{44}<L_{X}<10^{44.5}$ and the two slowly declining segments
between $10^{42.5}<L_{X}<10^{44}$ and $10^{45.5}<L_{X}<10^{46}$.
Comparing with the observed type-II fractions, the dusty torus
opening angle is inferred as around 70 degrees, in agreement with
the range of inclination angles determined for some Seyfert-I AGNs
\cite{Wu}. The predicted nearly 100\% type-II AGNs for
$L_{X}<10^{42}$ is the direct consequence of the intrinsic
luminosity function cutoff below 10$^{42.75}$ erg/s. Similarly the
predicted rapid decreases of type-II AGNs for
$10^{44}<L_{X}<10^{44.5}$ and $L_{X}>10^{46}$ are due to the
intrinsic luminosity function break and cutoff at $L_{X}\sim
10^{45}$ and $L_{X}>10^{46}$ respectively. Future AGN surveys with
more statistics for both low and high luminosity ends will test
the predictions of our model and thus measure the intrinsic AGN
luminosity function more accurately.

\section{Discussion and conclusion}
We first stress the point that because of the inclination angle
effects, the observed apparent luminosity of each AGN is not the
intrinsic luminosity of the AGN, unless the inclination angle of
each AGN is measured directly and the inclination angle effects
are corrected to recover the intrinsic luminosity for each AGN.
Lacking of inclination angle information for most AGNs, the
intrinsic luminosity function of AGNs is currently not observed
directly, because the observed apparent luminosity function is
already convolved with the inclination angle effects. We therefore
assumed a simple broken-power law form of the intrinsic luminosity
function and determined the parameter values by fitting the
convolved luminosity function (with $f(x)$) with the observed
apparent luminosity function. The functional form is not motivated
astrophysically, but simply chosen to obtain a good fit with the
observed apparent luminosity function with a minimum number of
free parameters. The good agreement between this simple form of
intrinsic luminosity function suggests that any reasonable AGN
synthesis model should be able to re-produce AGN intrinsic
luminosity function similar to that shown in Figure 1.

In this AGN unification model, we explicitly require that the
X-ray emission is mainly produced from an optically thick
accretion disk coaxed with the torus. For the typical type-I AGN
NGC~4151, its hard X-ray power-law exhibits a characteristic
cutoff above around 50 keV, which may be explained as due thermal
Comptonization of cold disk photons in a hot medium
\cite{Zdziarski}. Detailed modeling of the hard X-ray spectrum
resulted in a Comptonization $y$-parameter of
0.88$^{+0.12}_{-0.11}$ and an electron temperature of
73$^{+34}_{-29}$\cite{Zdziarski}, i.e., the Compton scattering
optical depth is 0.93-2.9, supporting our optically thick
assumption of the scattering medium.

Many observations are also consistent with the disk origin of AGN
X-ray emission. For example, the comparison between the
variabilities in the X-ray light curves of AGNs, intermediate mass
BH systems and X-ray BH binaries shows that the variability
timescales are proportional to their BH masses
\cite{Edelson,Lee,Vaughan,Strohmayer,Markowitz,Cropper}. This
demonstrates the same accretion disk origin of X-ray radiation
from all these systems, and thus similar physical processes may be
going on in astrophysical systems with entirely different scales
\cite{Zhang2}. In particular for the SMBH in the center of the
milky way, several disk oscillation modes are identified from its
X-ray flares \cite{Baganoff} which allowed very precise estimate
of the mass and angular momentum of the BH \cite{Aschenbach}. The
inverse Compton scattering process in the accretion disk may be
responsible for the observed X-ray emission \cite{Liu}.
Alternatively magnetic energy release may be responsible for X-ray
emissions from the solar and stellar coronae, and accretion disks
in X-ray binaries, intermediate mass and SMBHs, because in all
these systems X-ray flares are commonly seen \cite{Liu2}. A
disk-like patchy corona \cite{Haardt} in AGN disk may produce the
observed power-law like X-ray emission through magnetic
reconnection process \cite{Wang3}. Socrates, Davis \& Blaes (2004)
have pointed out recently that in the innermost regions of
radiation pressure supported accretion disks around black holes in
both stellar mass and supermassive black holes, the turbulent
magnetic pressure may greatly exceed the gas pressure.
Consequently turbulent Comptonization may be able to produce X-ray
photons in these accretion disks independent of the central black
hole mass, providing a viable mechanism for X-ray photon
production in AGN disks.

The assumption of the disk-torus alignment, as assumed previously
\cite{Wu}, is also natural. The formation of an accretion disk
requires significant amount of angular momentum for the material
transferred to the disk at larger radii. The only known structure
in an AGN immediately outside the accretion disk is the dusty
torus. Therefore the accretion disk and the torus should be
aligned if the torus is the source of the material forming the
accretion disk \cite{Krolik}.

We conclude that the AGN unification proposed about two decades
ago has survived another critical test. The success of our simple
model, in predicting the observed apparent X-ray luminosity of
AGNs and the type-II AGN fractions, calls for a unification model
for AGNs including a torus, an X-ray emitting accretion disk and a
central SMBH; we call this {\it ``TAXI"} model, which stands for
{\it T}orus of {\it A}ntonucci with {\it X}-ray {\it
I}nclination-angle effects. The inferred single intrinsic
luminosity function for AGNs, which is significantly different
from the observed apparent luminosity function, should be used in
the future for all AGN population synthesis and related studies.
Within the framework of this model, it is important to investigate
further the physics for the formation of the torus and its
relationship with the X-ray emitting accretion disk, in order to
understand the formation and evolution of SMBHs and galaxies
\cite{Kauffmann,Page,Menci}, which are intimately related to the
properties of dark matter and the evolutionary history of the
universe \cite{Baes,Matteo}.

\noindent {\bf Acknowledgement: }The anonymous referee is
appreciated for his/her comments and suggestions, which allowed us
to clarify some issues and improve the presentation of the paper
significantly. We thank Dr. Jian-Min Wang and Mr. Xin-Lin Zhou of
IHEP/CAS (China) for interesting discussions and many helpful
suggestions. We also thank the organizers and participants of the
2004 Annual IHEP-Tsinghua Student Astrophysics Symposium for many
stimulating discussions, which motivated us for this work. NSFC,
CAS and MOST in China and NASA in USA are acknowledged for partial
financial support to SNZ through several research grants.
\bibliographystyle{science}

\end{document}